\documentclass[preprint,showpacs,preprintnumbers,amsmath,amssymb,nofootinbib]{revtex4}
\usepackage{graphicx}
\usepackage{dcolumn}
\usepackage{bm}
\pdfoutput=1

\begin{document}
\title{\bf The Hilbert Lagrangian and Isometric Embedding:  Tetrad Formulation of Regge-Teitelboim Gravity}

\author{Frank B. Estabrook}

\affiliation{Jet Propulsion Laboratory, California Institute of Technology, Pasadena, CA 91109}

\date{\today}

\begin{abstract}We discuss Exterior Differential Systems (EDS) for the vacuum gravitational field.  These EDS are derived by varying the Hilbert-Einstein Lagrangian, given most elegantly as a Cartan 4-forrm calibrating 4-spaces embedded in ten flat dimensions.  In particular we thus formulate with tetrad equations the Regge-Teitelboim dynamics ``{\it \`{a} la} string" (R-T); it arises when variation of the 4-spaces gives the Euler-Lagrange equations of a multicontact field theory. We calculate the Cartan character table of this EDS, showing the field equations to be well posed with no gauge freedom. The Hilbert Lagrangian as usually varied over just the intrinsic curvature structure of a 4-space yields only a subset of this dynamics, viz., solutions satisfying additional conditions constraining them to be Ricci-flat. In the static spherically symmetric case we present a new tetrad embedding in flat six dimensions, which allows reduction of the R-T field equations to a quadrature;  the Schwarzschild metric is a special case.  As has previously been noted there may be a classical correspondence of R-T theory with the hidden dimensions of brane theory, and perhaps this extended general relativistic dynamics holds in extreme circumstances where it can be interpreted as including a sort of dark or bulk energy, even though no term with a cosmological constant is included in the Lagrangian.  As a multicontact system, canonical quantization should be straightforward. 
\end{abstract}
\maketitle
\baselineskip= 14pt

\section{Introduction}Regge and Teitelboim \cite{Regge} and independently Pav\v{s}i\v{c} \cite{Pavsic} and, later, we \cite{ERW}, have introduced an extended field theory of vacuum gravity by considering the Ricci curvature scalar field induced, on a 4-space, by the embedding of the 4-space into a flat 10 dimensional pseudo-Euclidean space.  This is just the Hilbert-Einstein Lagrangian density: weighting the 4-volume it can be taken as an action for the embedding, and its extremization is a variational principle distinguishing 4-spaces that are solutions of the field equations of R-T gravity.  (For comparison, if just the 4-volume is taken unweighted the action gives the partial differential equations of minimal/maximal embedding.  In the words of R-T, these actions are {\it \`{a} la} string.)  Regge and Teitelboim treated the theory with general coordinates in the embedded and embedding spaces, and the resulting field equations are difficult.  When static spherically symmetric solutions were considered the Schwarzschild solution was shown to be a special case satisfying constraints, but the general solution had to be treated with perturbation methods \cite{Deser}.

The motive for this "stringy" approach to classical gravitation has been of course to go on to quantum fields;  there have been a few general papers \cite{Tapia} \cite{Karasik} but most subsequent work has considered only the cosmological solutions and their quantized versions \cite{Davidson} \cite{Davidson1} \cite{Cordero}.  

In \cite{ERW} we formulated R-T gravity using Cartan's method of moving frames, or tetrad formalism.  This technique of differential geometry has long been used for Einstein relativity and further is especially adapted to embedding problems. Use of anholonomic basis forms allow sets of partial differential equations to be algebraically analyzed as Exterior Differential Systems (EDS), calculating their Cartan characters.

Use of Cartan characters as a diagnostic tool for EDS is discussed, e. g., in Ivey and Landsberg \cite{Ivey}.  Given the generators of an EDS, 1-forms, 2-forms, 3-forms, and so on, one must calculate (at a generic point in the space of all variables) the ranks of a set of successively larger matrices for the construction of nested vectors spanning a local volume element that annuls the generators.  This is a very rapidly growing calculation, though the result, for the field theories of physics, the Cartan integer table, is often remarkably simple and powerful for insights into computer algorithms, conservation laws, constraints, boundary integrations.  We report in Section II a number of these tables for vacuum gravity, and in particular now for R-T theory. 

We will write the Cartan integer tables of our EDS' as $N(s_0,s_1,s_2,s_3)4 + CC$.  N is the total number of field variables, dependent and independent. $CC$ is the number of Cauchy Characteristic vectors in a solution; they arise from the absence of that number of basis 1-forms in the generators of the EDS, so in principle that many fields can be completely eliminated.  $s_i  (i=0,1,2,3)$ are the calculated Cartan integers, and $s_4$ is defined to be $N-CC-4-s_0-s_1-s_2-s_3$.  It is the number of gauge degrees of freedom.

The total number of first order partial differential equations coded by an EDS is $4s_0 + 3s_1 +2 s_2 +s_3$; of these $s_0 + s_1 + s_2 +s_3$ can be taken as evolution equations in a timelike direction, there are $s_0 + s_1 + s_2 $ so-called first class constraints, $s_0 + s_1  $ second class constraints and $s_0$ third class.  Together of course the constraint equations are the integrability conditions computable from commutators of the evolution equations.

To calculate the Cartan characters, we have used a suite of Mathematica programs written by H. D. Wahlquist \cite{WAhl}.  It is important that his program characters.nb further checks that a specified set of four basis forms remain linearly independent (or in Cartan terminology, ``in involution") in the four dimensional solutions.  In the following it is always the 1-forms $\theta ^i$ that have been explicitly confirmed to be independent, so becoming an orthonormal tetrad field in the embedded solutions.

In Sections III, IV, and V we specialize the  R-T embedding EDS to the static spherically symmetric case.  Just six flat dimensions suffices;  we first find a new embedding of tetrads consistent with the known coordinate embedding in this case.  It results in a manageable new set of pde's for the R-T theory, which we reduce to a quadrature and can plot.  The specialization to obtain the Schwarzschild metric become clear, together with its neighboring solutions.

In Section VI for completeness we formulate the spatially homogeneous isotropic solutions of embedding class one;  this case of R-T gravitation has previously been considered by several authors, with the primary interest of going on to quantum cosmology {\it \`{a} la} string.

\section{First Order Lagrangians and External Differential Systems for General Relativity}The mathematical structure of the coordinate version of tetrad equations for vacuum general relativity has been explored using Cartan theory, formulating those equations as an Exterior Differential System (EDS) \cite{FEMath}. That tetrad EDS is equivalent to Newman-Penrose or dyadic sets of first order partial differential equations.  The EDS follows from a Cartan 4-form and its exterior derivative the Cartan-Poincar\'{e}, or multisymplectic 5-form, the terms of which factor into products of torsion 2-forms and Einstein 3-forms that generate the EDS. The Cartan characters of the EDS, and resulting proof of well posedness, were calculated for that case, where there were 40 dependent variables (16 coordinate components of the tetrads, and 24 tetrad components of the connection) and 4 independent variables $x^{i}$. The Cartan character table is 44(0,4,12,14)4;  from these Cartan theory shows that solutions are well posed from boundary conditions, and have 10 degrees of gauge freedom ($s_{4}=10$). There are 30 evolution equations and 20 constraint equations.  There are also 10 invariance generators which together with the Cartan form lead to 10 conservation laws \cite{FECons}.  The first order functional Lagrangian equivalent to this Cartan form has been repeatedly treated, in work going back to M{\o}ller and Weyl.  The variational formulation of vacuum Maxwell and Yang-Mills equations in p dimensions parallels this:  the multisymplectic forms are sums of terms each of which is a product of  a 2-form and a p-1-form \cite{FEYang}.  These EDS all are there given directly in terms of scalar variables, dependent fields and coordinates. 

A more geometric and elegant EDS would have the property of ``constant coefficients", being expressed in terms only of sets of (non-exact) basis 1-forms with constant coefficients, and so with no explicit coordinate dependence.  The structure relations of the basis forms then are separately given, demonstrating an underlying Lie group algebra.  The Cartan characters and Lie derivatives of such a ``cc" EDS are readily computable at any point, and properties they code such as numbers of evolution and constraint equations, associated conservation laws and potentials, much more easily and generally obtained.  Of course to actually obtain solutions of a ``cc" EDS one must finally re-introduce explicit scalar coordinates and use exact basis forms while respecting the Lie structure.   

For vacuum general relativity the Hilbert Lagrangian, expressed as a Cartan 4-form, yields a ``cc" formulation when used to minimize the induced Ricci-scalar action of a metric 4-space embedded, with its orthornormal frame bundle, into that of a flat 10-space \cite{ERW}. Which is to say the use of non-exact basis 1-forms from the Cartan-Maurer equations of the O(10) Lie group in fibers over 10-space. This bundle is 55 dimensional. Several families of ``cc" EDS's of this sort have been discussed in detail in a recent summary paper \cite{FEGIFT}.  Large scale topological questions aside, it is well known that 10 dimensions are sufficient for the local embedding space of general four dimensional curved spaces. With special symmetries the required flat embedding space may be of smaller dimension (embedding classes 1-6 \cite{Exact}).

The exterior derivative of the Hilbert Cartan 4-form for ``cc" embedding in flat 10-space, the multisymplectic 5-form, can be written as a sum of four terms, each of which is a product of a torsion 2-form and an Einstein 3-form in exact parallel to the above tetrad-derived multisymplectic form. The exterior derivatives of the torsion 2-forms must be included for closure, and together they then generate an EDS whose Cartan character table is almost identical to that of the coordinate version of tetrad vacuum relativity, 55(0,4,12,14)4 + 21CC but now there is no gauge freedom ($s_{4}=0$), since there are 21 Cauchy Characteristic vectors, or ``CC" \cite{Ivey}, which implies that in fact 21 variables and basis 1-forms could at least in principle be completely eliminated, and the EDS set on a space of only 34 dimensions. These Cauchy vectors of the EDS obviously arise from arbitrary Lorentz transformation of the 4-frames at each point of a 4-dimensional solution, and from arbitrary six dimensional orthogonal transformation in the coframe space there:  6 + 15 connection basis forms corresponding to these are in fact absent from the Cartan form. 

If such an EDS for vacuum gravity, generated by torsion 2-forms and Einstein 3-forms, is set on an embedding space of less than ten dimensions the Cartan character table shows it NOT to be well posed.

The embedding multisymplectic 5-form is however writeable in a second way, to a sum of six terms that have ``contact" 1-form and 4-form factors.  This in fact might be claimed to be the most fundamental way in which an EDS should arise from a field theoretic variational principle, generalizing the simplectic geometry of classical Hamiltonian particle dynamics. There is a extended body of work on this by Betounes, Bryant, Carath\'{e}odory, Dedecker, DeDonder, Gotay, Griffiths, Hermann, Isenberg, Kijowski, Krupka and Krupkov\'{a}, Lepage,  Marsden, Montgomery, Olver, Rund, Shadwick, \'{S}niatycki, Sternberg, Szczyrba, Tulczyew, and a number of others.  A rigorous differential geometric  review of this ``multi-contact" approach and the Euler-Lagrange equations that result, by Bryant et al, is recommended \cite{Bry}. For fiber bundle formulations, and other references to the multisymplectic formulation of relativistic field theories cf. Gotay et al \cite{Gotay}.  When the contact 1-forms that vanish on solutions of such an EDS are basis forms of an orthonormal frame bundle, the system is properly termed an isometric embedding.  The Cartan character table of the isometric embedding EDS arising from the Hilbert Lagrangian for four dimensions embedded in ten is 55(6,6,6,12)4 + 21CC.  This is R-T field theory, clearly more general than the Einstein theory coded in the previous ``cc" EDS. The latter EDS is included in the former, so all solutions of the former EDS are included among those of the latter.  Similar well posed character tables are found for isometric embedding in all other dimensions, in particular for embedding classes 1-5.             

Finally, if the 3-form generators of the previous theory are added to the isometric 1-form and 4-form generators, yet another ``cc" EDS results with 55(6,6,10,8)4 + 21CC, now again constraining the solutions to be Ricci-flat \cite{ERW}.  This character table shows four first class constraints to have been added, which was in fact done at the end of Regge-Teitelboim;  they conjectured that these equations for Einstein vacuum theory were consistent and this table indeed shows it to be well-posed. Isometric vacuum EDS' for embedding classes 1-5 are not well posed; embedded in eleven or more dimensions they have gauge freedom.

\section{Embedding of Spherically Symmetric Static Geometries into Six Dimensions}  It is known that a general spherically symmetric four dimensional metric is of embedding class two, i.e, requiring a minimum of six flat dimensions \cite{Kar}\cite{Exact}.  So we now specialize to spherically symmetric static 4-metrics isometrically embedded in flat six.  We assume that the Euler-Lagrange equations still capture all the essential dynamics with this imposed symmetry. The structure equations of the 21 basis forms spanning the orthonormal frame bundle over flat 6-space are
\begin{eqnarray}
{d\theta }^{\mu}+{\omega^{\mu}}_{ \nu} \wedge {\theta }^{\nu } &=& 0 \\
{d\omega^{\mu }}_{\nu }+{\omega ^{\mu }}_{\sigma }\wedge {\omega
^{\sigma }}_{\nu }
 &=& 0.
\end{eqnarray}
$\mu$ and $\nu$ range from 1 to 6; we will later lower/raise the hexad indices with an appropriate diagonal signature matrix.  In this moving frame formalism, in any cross section the $\theta^{\mu}$ 1-forms are an orthnormal hexad, and the $\omega^{\mu}_{\nu}=-\omega^{\nu}_{\mu}$ the connection 1-forms.  Expanding them on the $\theta^{\mu}$ basis defines the 
Ricci rotation coefficients of the hexad field: $\omega_{\mu \nu }=\Gamma_{\sigma \nu \mu} \theta ^{\sigma}$.

These equations can be rewritten with indices having ranges i,j = 1...4 and A,B = 5,6 that will be used in the isometric embedding:  

\begin{eqnarray}
d{\theta ^i}+{\omega ^i}_j \wedge {\theta ^j} &=& -{\omega ^i}_A
\wedge {\theta ^A}\\
{d\theta ^A}+{\omega ^A}_B \wedge {\theta ^B} &=& -{\omega ^A}_i
\wedge {\theta ^i} \\
{d\omega ^i}_j+{\omega ^i}_k \wedge {\omega ^k}_j &=& - {\omega ^i}_A
\wedge
{\omega ^A}_j \\
{d\omega ^A}_B +{\omega ^A}_C \wedge {\omega ^C}_B &=&-{\omega ^A}_i
\wedge
{\omega ^i}_B  \\
{d\omega ^i}_A +{\omega ^i}_j \wedge {\omega ^j}_A + 
{\omega ^i}_B \wedge {\omega ^B}_A &=& 0.
\end{eqnarray}

The terms we have put on the right are the torsions and Riemann curvatures induced by an embedding $\theta^A=0$; we will use them to set the Cartan Form and EDS.  The former is (now assuming a signature matrix is being used)
\begin{equation}
\Lambda  = {R_{{ij}}}\wedge {{\theta }_k}\wedge {{\theta }_l}
{{\epsilon }^{{ijkl}}},
\end{equation}
where from the Gauss Equation (5) $2{R_{{ij}}} :=
-{{\omega }_{{iA}}}\wedge \omega_{j}^{A}$
is the induced Riemann 2-form.  In any solution, an embedded 4-space, this Cartan form pulls back to be the Ricci scalar, but it is its functional form, implied from taking it a a calibration on arbitrary 4-volume elements in the embedding space, that generates the variational formulation of the field theory.  The exterior derivative of the 4-form
field $\Lambda$ is quickly
calculated using Equations (3) and (7) to be the multisymplectic 5-form
\begin{equation}
d \Lambda = \theta^{A} \wedge \omega_{Ai} \wedge {R_{jk}}\wedge \theta
_{l} \epsilon^{ijkl}.
\end{equation}
This is factored term by term to generate an embedding EDS.  (In this geometric formulation factorization is the essence of a variational principle, for then an arbitrary deformation vector field, acting as a Lie derivative on $\Lambda$, always gives 4-forms that are in the ideal of the EDS.)  The first ``cc" EDS we discussed in Section II was generated by four 2-forms $\omega_{Ai}\wedge \theta^A$ and four 3-forms $R_{jk} \wedge \theta_l \epsilon^{ijkl}$, plus their exterior derivatives for closure. It required an embedding space of 10 dimensions to be well posed.  The more general isometric EDS is generated by 1-form and 4-form factors $\theta^A$ and  $\omega_{Ai} \wedge 
R_{jk} \wedge \theta_l \epsilon^{ijkl}$, plus the 2-form exterior derivatives of the 1-forms, for closure, $\omega^{A}_{i} \wedge \theta^i$. For an embedding space of six dimensions the character table is 21(2,2,2,4)4 + 7CC. 

Solving this EDS now requires using the remaining structure Equations (4) and (6) to find explicit scalar variables, which will become dependent and independent variables in a set of partial differential equations, and using their exterior derivatives as 1-form bases. Happily in this specialized case perhaps half the work has been done, as we have a coordinate expression for the class two embedding of general spherically symmetric static 4-dimensional metrics into 6-dimensional flat space already worked out by Ikeda et al \cite{IKEDA}. We have gone farther using the structure equations, or equivalently orthonormalization to construct a coordinate expression of an adapted "hexad" frame -- a set of six orthonormal basis 1-forms $\theta^{\mu}$  -- that pulls back to an othornormal tetrad field in the embedded 4-metric. We adopt one of the three possible signatures considered by Ikeda et al for the metric of the embedding flat space: (-,-,-,+,+,-).  The first four signatures will become those of embedded space-time. This orthonormal moving frame adapted to spherically symmetric embedding is next given in terms of Ikeda et al's pseudo-Euclidean coordinates $r, \theta, \phi, z^4, z^5, z^6$. It depends on just two unknown functions of the radial coordinate $r$, denoted $c[r]$ and $g'[r]$, which are essentially the two arbitrary functions used by Ikeda et al.  We also write $t$ = $z^4/2c[r]$.
\begin{eqnarray}
\theta_1 =(dr + (dz^5 + dz^6) c'[r]-2 c'[r] t (2 c'[r] t dr + 2 c[r] dt) \nonumber \\ 
 -(dz^5 - dz^6)(g'[r] + c'[r] t^2))/\sqrt{1 +4 c'[r] g'[r]}
\end{eqnarray}
\begin{equation}
\theta_2 = r d\theta, \;  \; \theta_3 = 
 r \sin[\theta] d\phi,  \; \; \theta_4 = 
 t dz^5 - t dz^6 + 2 c[r] dt + 2 c'[r] t dr
\end{equation}
\begin{eqnarray}
\theta_5 = (dr + dz^5 (g'[r] - c'[r] t^2 + c'[r] + 1/(2 c'[r]))- 2 t c'[r] (2 c'[r] t dr + 2 c[r] dt) \nonumber \\
    - dz^6 (g'[r] - c'[r] t^2 - c'[r] + 1/(2 c'[r])))/\sqrt{1 +4 c'[r] g'[r]}
\end{eqnarray}
\begin{equation}
\theta_6 = dr + dz^5/(2 c'[r]) - dz^6/(2 c'[r])
\end{equation}
We entered this orthonormal framing into the Mathematica program AVF.nb \cite{WAhl} which then calculates all the rotation coefficients, connection forms $\omega_{\mu \nu}$, and Riemann, Ricci and Weyl tensor components of a given orthonormal frame, here in terms of the functions $c[r]$ and $g'[r]$.  It calculates zero for all 105 components of the Riemann tensor, which verifies that indeed we have a flat embedding space. The 1-form isometric embedding equations are $\theta^A=0$, for $A=5,6$.  Evaluating these allows solution for $dz^5$ and $dz^6$.  The pullback of the 6-metric to the embedded space is 
\begin{equation}-\theta^1 \theta^1-\theta^2 \theta^2-\theta^3 \theta^3+\theta^4 \theta^4
\end{equation}  
Evaluating this and substituting $dz^5$ and $dz^6$ we obtain Ikeda et al's spherically symmetric static line element:
\begin{equation}
ds^2= -(1 + 4 c'[r] g'[r]) dr^2 - r^2 d\theta^2  -  r^2 sin^2[\theta] d\phi^2 + 4 c[r]^2 dt^2 
\end{equation}

We then calculated the closure 2-forms in the EDS, $\omega_{Ai}\wedge\theta ^i$, and find that each term involves $\theta^5$ or $\theta^6$ so these 2-forms add no equations. This automatic lack of torsion in the solution stems from the use of tetrads tied to an (Ikeda et al) coordinate frame.

Two Euler-Lagrange equations from the 4-forms  ${{\omega }_{{Ai}}}\wedge
{R_{{jk}}}\wedge {{\theta
}_l} {{\epsilon }^{{ijkl}}}$ remain for solution;  the Mathematica program calculates each in the form of a function times $\theta^1 \wedge \theta^2 \wedge \theta^3 \wedge \theta^4$.  Their vanishing results in dynamic equations that are quite nonlinear in the unknowns $c[r]$ and $g'[r]$ and their derivatives $c'[r]$, $c''[r]$ and $g''[r]$.

\section{Dynamics}We first make the following convenient change of variables, to $s[r]$ and $\mu[r]$;
\begin{equation}
g'[r] = r \mu[r]/(2 c[r]), \; c'[r] = s[r] c[r] /r
\end{equation}
We find that c[r] drops out and we then have two quasi-linear first order ordinary differential equations that can be solved for the first derivatives of  $s[r]$ and $\mu[r]$.  This is the dynamics we must explore:
\begin{equation}
r s'[r]=s[r]\frac{(2 -4s[r]-6 \mu[r]+3s[r] \mu[r]-2s[r]^2 \mu[r]+3\mu[r]^2-3s[r]\mu[r]^2-2 s[r]^2\mu[r]^2)}{(4-6\mu[r]+2s[r]\mu[r]+3\mu[r]^2)}
\end{equation}
\begin{equation}
r \mu'[r]=\mu[r](1- \mu)\frac{(-6 +4s[r]+3 \mu[r]-9s[r] \mu[r]+2s[r]^2\mu[r])}{(4-6\mu[r]+2s[r]\mu[r]+3\mu[r]^2)}
\end{equation}
Two integration constants would arise if we could find a closed form general solution.  The equations are invariant under scale change of $r$, so one may use $r/2m$ and indeed reduce them to a single ordinary differential equation (next Section).  A third integration constant arises in the quadrature for $c[r]$, Equation (16), but this can be absorbed into the coordinate $t$.

The dynamical Equations (17) and (18) can now be entered in the expressions Mathematica calculated for the induced Riemann tensor in Equation (5). The non-vanishing tetrad components of the Einstein tensor of solutions of this extended theory are:
\begin{eqnarray}
E_{11}=-\frac{2 s[r] (1 -\mu[r])}{r^2 (1 + 2 s[r] \mu[r])}
\end{eqnarray}
\begin{eqnarray}
E_{22}= E_{33}=-\frac{s[r] (1 - \mu[r]) (2 + s[r] \mu[r])}{r^2 (1 + 2 s[r] \mu[r]) (4 + 2 (-3 + s[r]) \mu[r] + 3 \mu[r]^2)}
\end{eqnarray}
\begin{eqnarray}
E_{44}=\frac{6 s[r] \mu[r]^2) (1 - \mu[r])}{r^2 (1 + 2 s[r] \mu[r]) (4 + 2 (-3 + s[r])\mu[r] + 3\mu[r]^2)}
\end{eqnarray}
The trace, or negative Ricci scalar, is
\begin{eqnarray}
TrE = -\frac{6 s[r] (1 - \mu[r]) (2 + (-2 + s[r]) \mu[r] + 2 \mu[r]^2)}{r^2 (1 + 2 s[r] \mu[r]) (4 + 2 (-3 + s[r]) \mu[r] + 3 \mu[r]^2)}
\end{eqnarray}
The Weyl tensor is conveniently written as traceless symmetric dyadic 3x3 arrays of "electric" components $A_{ab}$ and "magnetic" components $B_{ab}$ a,b = 1,2,3 in the $\theta^i$ frame.  We find Petrov type D, the $B_{ab}$ vanishing and $A_{ab}$ diagonal:
\begin{equation}-A_{11}=2A_{22}=2A_{33}=\frac{s[r]}{r^2}\frac{(2 + (2 + s[r]) \mu[r] + (-4 + 3 s[r]) \mu[r]^2 + 2 \mu[r]^3)}{(1 + 2 s[r] \mu[r]) (4 + 2 (-3 + s[r]) \mu[r] + 3 \mu[r]^2)}
\end{equation}
Remember, in terms of variables $\mu[r]$ and $s[r]$ the line element Equation (15) has become
\begin{equation}
ds^2= -(1 + 2 s[r] \mu[r]) dr^2 - r^2 d\theta^2  -  r^2 sin^2[\theta] d\phi^2 + 4 c[r]^2 dt^2
\end{equation}
where $c[r]=\exp (\int(s[r]/r)dr$).
\begin{figure}
\includegraphics[width=5in]{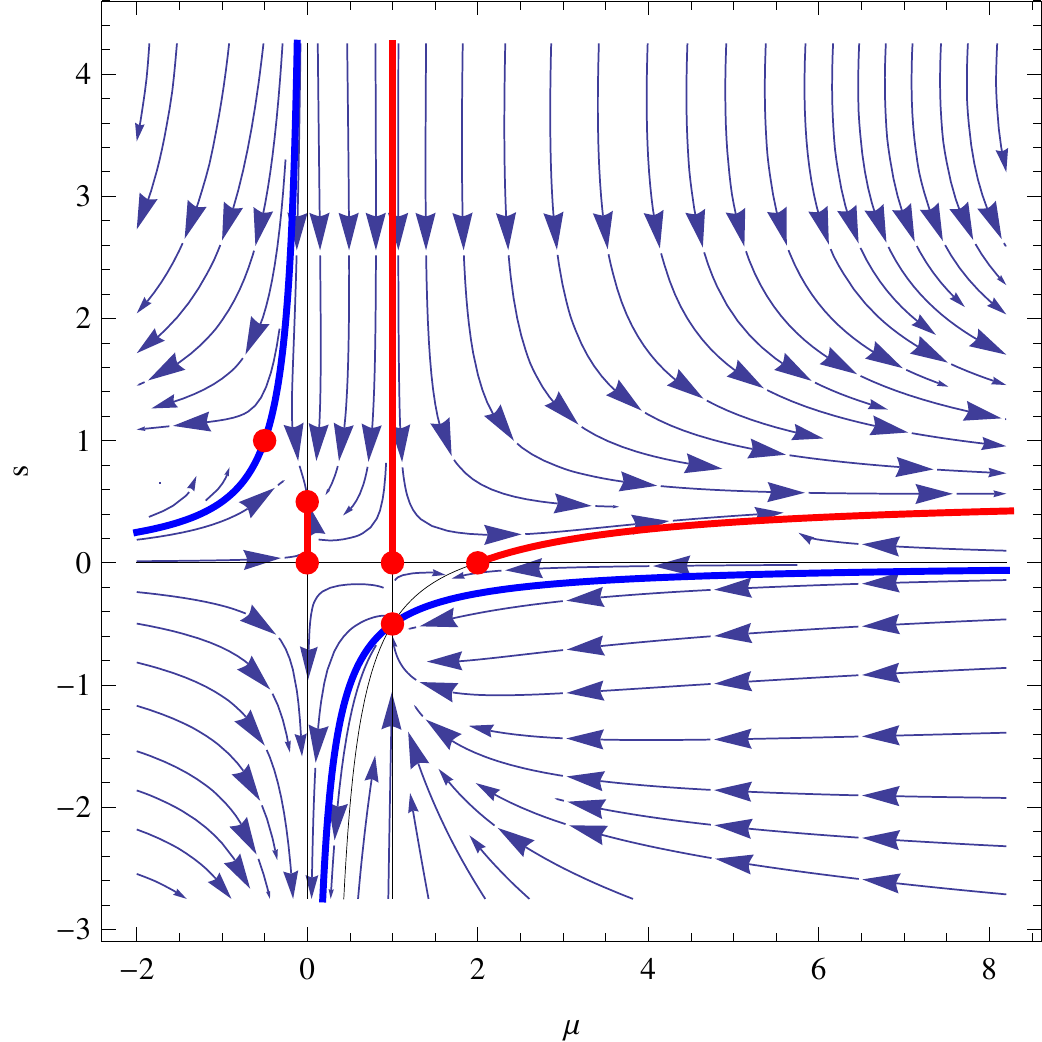}
\caption{$ds/d\mu$ flow.  A one parameter family of trajectories, each of which is a spherically symmetric static solution of the extended Hilbert dynamics.  The Schwarzschild solution is the red trajectory along $\mu = 1$.\label{fig:flow}}
\end{figure}

\section{Flow Plot of Solutions}The independent variable $r$ enters the equations for $s'[r]$ and $\mu'[r]$ in an autonomous way, so solutions can be classified by projecting to a vector field in $s$, $\mu$ space:  
\begin{equation}
ds/d\mu=(s/\mu)\frac{2 -4s-6 \mu+3s \mu-2s^2\mu+3\mu^2-3s\mu^2 -2 s^2\mu^2}{(1-\mu)(-6 +3\mu+4s -9\mu s+2s^2\mu)}
\end{equation}
We have not found a general closed form solution, but a numerically integrated ``flow" plot is illuminating (Figure 1). The polynomials in the numerator and denominator of Equation (25) have six common roots, which are shown in the Figure as red dots. These are limit points for integration of the dynamical system. Our solutions are trajectories between these dots, or from them outward, each parametrized by $r/2m$, running from 0 to infinity.  The two trajectories shown in heavier blue are where the metric of Equation (15) or (24) is singular: $1 + 4 c'[r] g'[r]=1 + 2 s[r] \mu[r]=0$, so we are mainly interested in red dots and trajectories lying between them. The point at $(\mu,s)=(0,1/2)$ is an attractor; the other three are saddle points.  There is also an attractor at $\mu = \infty$, $s = (\sqrt{33}-3)/4$. Equations (19) to (23) show trajectories along $s = 0$ to be flat 4-spaces.  We will only discuss the limiting solutions corresponding to the three trajectories shown in red.

The Schwarzschild metric is the trajectory along $\mu =1$ from $s = \infty$ to the saddle point at $s = 0$, the Einstein tensor vanishing on it.  We find
\begin{equation}s[r]=\frac{m/r}{1-2m/r} 
\end{equation}
\begin{equation}
4 c[r]^2=1-2m/r
\end{equation}
From Equation (23) the Weyl component $A_{11}=-2m/r^3$.  This solution however appears to be unstable to slight variations of initial conditions, as closely adjacent trajectories veer toward the two attractors. These varied solutions in fact are variations of the other two red trajectories, which are stable, and so perhaps their metrics are of interest for applications of the extended dynamics.  We consider them briefly.

The trajectory along $\mu =0$ from $s=0$ to $s=1/2$ can be as quickly integrated to give 
\begin{eqnarray}s[r]=\frac{r/2m}{r/m+\sqrt{r/2m}}\\c[r]=1+2\sqrt{r/2m}
\end{eqnarray}which can be substituted into the Einstein and Weyl components, and into the metric Equation (24).  These fields all behave as $r^{-2}$ at large $r$.

We have not found a closed expression for the trajectory beginning at $s=0$, $\mu = 2$, extending to $s = (\sqrt{33}-3)/4$ at $\mu = \infty$, but it is immediate that on it again the Einstein and Weyl fields behave as $r^{-2}$ at large $r$.  

\section{Cosmology}The embedding class of 4-metrics having an invariance group $G6$ of rotations and translations, acting on three dimensional subspaces, is one. Embedding coordinates are known \cite{Exact}, and orthonormal basis forms then easy to find; we give those appropriate to solutions having hyperbolic 3-dimensional subspaces:
\begin{equation}
\theta _1=
   \text{d$\chi $} R(t), \; \; \theta
   _2= \text{d$\theta $} R(t)
   \sinh (\chi ), \; \; \theta _3=
   -\text{d$\phi $} R(t)
   \sinh (\chi ) \sin (\theta
   )\nonumber\end{equation}
   \begin{equation}
   \theta _4= \text{dy}
   \sqrt{R'(t)^2-1}-\text{dt}
   R'(t)^2,\; \; \theta _5=
   \text{dy} R'(t)-\text{dt}
   R'(t)
   \sqrt{R'(t)^2-1}
\end{equation}
The coordinates we use in this 5-space are $\chi, \theta, \phi, t$ and $y$.  $R(t)$ is an arbitrary function characterizing the subset of coordinate systems and pentad frames which respect the symmetry group.  The signature for these basis 1-forms is (-1,-1,-1,+1,-1).  The metric of the 5-space is flat:
\begin{eqnarray}ds^2=-\theta^1 \theta^1-\theta^2 \theta^2-\theta^3 \theta^3+\theta^4 \theta^4-\theta^5 \theta^5=\nonumber \\\text{dt}^2
   R'(t)^2-\text{dy}^2-\text{d$\phi $}^2 R(t)^2 \sinh
   ^2(\chi ) \sin ^2(\theta
   )-\text{d$\chi $}^2
   R(t)^2-\text{d$\theta $}^2
   R(t)^2 \sinh ^2(\chi )
\end{eqnarray}
From these the AVF program then calculates, as above, Ricci Rotation Coefficients $\Gamma_{\mu \nu \sigma}$, the connection 1-forms $\omega_{\mu \nu}$ and Riemann 2-forms $R_{\mu \nu}$.  The last all vanish.

An isometric embedding first sets $\theta_5$ to zero, allowing solution for $dy$ in terms of the other four coordinate bases.  Equations (8), (9) and (14) above apply as given with A = 5. Pulling back the 5-metric, and substituting for dy, gives the metric of the embedded 4-space
\begin{equation}
ds^2=-\text{dt}^2+\text{d$\phi
   $}^2 R(t)^2 \sinh ^2(\chi )
   \sin ^2(\theta )+\text{d$\chi
   $}^2 R(t)^2+\text{d$\theta
   $}^2 R(t)^2 \sinh ^2(\chi )
   \end{equation}
Tetrad components of the induced Einstein tensor of the embedded 4-space, computed from the $\Gamma_{ijk}$, are those of a perfect fluid
\begin{equation}
E_{11}=E_{22}=E_{33}=(1-\dot{R}(t)^2-2 R(t)\ddot{R}(t))/R(t)^2
\end{equation}
\begin{equation}
E_{44}=3(-1+\dot{R}(t)^2)/R(t)^2
\end{equation}

Pulling back the single dynamic 4-form from the Hilbert variational EDS yields a second order equation for the kinematics of these models
\begin{equation}
-1+\dot{R}(t)^2 + 3 R(t) \ddot{R}(t)=0
\end{equation}
and this integrates to give an analog to the Friedmann equation
\begin{equation}
\dot{R}(t)^2=1+ K R(t)^{-2/3}
\end{equation}
K is an arbitrary constant.  Taking it positive, and substituting into Equation (33) and (34) gives expanding decelerating solutions with negative pressure
\begin{equation}
E_{11}=E_{22}=E_{33}=p/c^2= -\frac{K}{3 R(t)^{8/3}}
\end{equation}
\begin{equation}
E_{44}=\rho=\frac{3K}{R(t)^{8/3}}
\end{equation}

\section{Summary and Comments}We have formulated an extended geometric theory of gravity using Cartan's method of moving frames in 10 flat dimensions.  This R-T theory is a well posed field theory, now coded as a multicontact Exterior Differential System generated by 1-forms for isometric embedding of four-dimensional Riemannian geometry and by 4-forms completing the Euler-Lagrange equations for the first order Hilbert action.  It is of course well known that this action, when varied over the intrinsic curvature of a 4-geometry, yields Einstein's vacuum field equations:  Ricci-flat general relativity.  One should note however that there is then a degeneracy in that the value of the Lagrangian action, the Ricci scalar, vanishes on solutions.  When the same Lagrangian is taken as a variational principle {\it \`{a} la} string extra freedom appears and Ricci-flat solutions are now special cases, found only with particular initial conditions for evolution of the field dynamics.

Whether the new more general solutions of this extended theory have immediate physical meaning is not yet clear, but mathematically this is a multicontact dynamical system, a field theory correctly generalizing classical Hamiltonian particle dynamics.  Its canonical quantization should consequently be straightforward.  It may be that the non-Ricci-flat solutions include a sort of dark energy, of purely gravitational origin, not arising from ad hoc addition of a cosmological constant but important in strong fields.

We worked out two highly symmetric cases as examples.  Imposing static spherical symmetry enables the variational problem to be reduced to six embedding dimensions, for which embedding coordinates are known, and a new orthonormal hexad frame for this was derived.  Partially integrating the tetrad Euler-Lagrange equations showed how the Ricci-flat (Schwarzschild) solutions are special.

Finally we worked out the "open cosmological" case  of G6 symmetry:  solutions containing nested hyperbolic 3-geometries.  Five embedding dimensions suffice.  A Friedmann-like equation was obtained for the timelike evolution of the spatial scale $R(t)$.  The gravitational field by intself acts like a cosmic fluid with negative pressure \cite{Davidson} \cite{Davidson1} \cite{Cordero}.
 
\section{Acknowledgements}This work was performed while I held a visiting appointment at the Jet Propulsion Laboratory.  My thanks to Dr. Moustafa T. Chahine and the Office of the Chief Scientist, and to the Science Division for hospitality.

\end{document}